\documentclass[aps,pra, twocolumn,superscriptaddress,amsfonts,a4paper,showpacs]{revtex4}
\usepackage{amsmath}
\usepackage{bm}
\usepackage{graphicx}
\def\BE {\begin{equation}}
\def\EE {\end{equation}}
\def\BEA {\begin{eqnarray}}
\def\EEA {\end{eqnarray}}
\def\BES {\begin{subequations}}
\def\EES {\end{subequations}}
\def\BA {\begin{array}}
\def\EA {\end{array}}
\def\NN {\nonumber}
\def\NN {\nonumber}
\def\ep {\varepsilon}

\begin{document}

\title{Quantum search by parallel eigenvalue adiabatic passage}
\author{D. Daems}
\affiliation{QuIC, Ecole Polytechnique,
Universit\'e Libre de Bruxelles, 1050 Bruxelles, Belgium}
\email{ddaems@ulb.ac.be}

\author{S. Gu\'erin}
\affiliation{Institut Carnot de Bourgogne UMR 5209 CNRS,
Universit\'e de Bourgogne, BP 47870, 21078 Dijon, France}
\email{sguerin@u-bourgogne.fr}

\author{N. J. Cerf}
\affiliation{QuIC, Ecole Polytechnique,
Universit\'e Libre de Bruxelles, 1050 Bruxelles, Belgium}

\abstract{We propose a strategy to achieve the Grover
search algorithm  by adiabatic passage in a very efficient way. An adiabatic process can be characterized by the instantaneous eigenvalues of the pertaining Hamiltonian, some of which form a gap. The key to
the efficiency is based on the use of parallel eigenvalues. This allows us to obtain non-adiabatic losses  which are exponentially small, independently of the number of items in the database in which the search is performed.
}}

 \maketitle

\section{Introduction}
Quantum computation by adiabatic evolution has been proposed as a
general method of solving search problems, mainly to exploit its
robustness towards unitary control errors and decoherence
\cite{adiabatic1,adiabatic2}. In contrast to the standard paradigm
of quantum computation \cite{nielsen}, which is implemented
through gates embedded in a quantum circuit, continuous-time
algorithms \cite{farhi}, and in particular adiabatic ones
\cite{adiabatic1,adiabatic2,RC}  proceed through the controlled
evolution of some Hamiltonians designed to solve the specified
problem. The adiabatic Grover algorithm, for instance, involves a
time-dependent Hamiltonian which smoothly drives the system, in a
time exhibiting a quadratic speedup, from one of its eigenstates
$|w\rangle$ that is easily prepared to a connected eigenstate that
coincides with the marked entry $|m\rangle$ of the database. This
can be achieved with the two-parameter Hamiltonian \BE
\label{model} H=a(t)H_\text{i}+b(t)H_\text{f}, \EE where
$H_\text{i}=|w\rangle \langle w|$ and $H_\text{f}=|m\rangle
\langle m|$ are simply projectors on the appropriate states while
$a$ and $b$ are time-dependent parameters which vanish at the
final and initial times, respectively, to ensure that the prepared
and target states are eigenstates.

The eigenstates of $H$ form in general an avoided crossing as a
function of time. The search is achieved when the dynamics follows
adiabatically the instantaneous eigenstate connected initially to
the prepared state $|w\rangle$ and finally to the marked state
$|m\rangle$. The way the parameters $a$ and $b$ vary around
the avoided crossing is key to making the search exhibit
or not a quadratic speedup. It has been shown that achieving a
quadratic speedup requires a non-linear dynamics of the
parameters: The dynamics has to slow down when approaching the
smallest gap of the avoided crossing and has to accelerate
afterwards \cite{RC}. This strategy will be referred to as
\textit{local strategy}.

In this paper we apply the strategy of optimal adiabatic passage
developed in Refs. \cite{Leveline,GJ} for two-level models of the form
(with real couplings)
\begin{equation}
H=\left[\begin{array}{cc} \Delta(t) & \Omega(t)\\
\Omega(t) & -\Delta(t)\end{array}\right].
\end{equation}
It has been shown that, for a given smooth pulsed-shape coupling
of the form $\Omega(t)=\Omega_0\Lambda(t)$ (with
$\Lambda(\pm\infty)=0$) and for the parametrization
$\Delta(t)/\Delta_0+\Omega(t)/\Omega_0=1$, the population transfer
is the most efficient, in the adiabatic regime, when the
instantaneous eigenvalues are \textit{parallel}. This corresponds
to $\Delta_0=\Omega_0$, i.e. to level lines (corresponding to
circles of equation $\Omega(t)^2+\Delta(t)^2=\Delta_0^2$) in the
diagram of the difference of the eigenvalue surfaces as a function
of the two parameters $\Omega(t)$ and $\Delta(t)$.

We show that this strategy, referred to as \textit{parallel
strategy}, applied to the problem of quantum search using the
two-parameter Hamiltonian (\ref{model}) leads to a Grover type
search, i.e. scaling with time as $\sqrt{N}$. It is moreover more
efficient than the local strategy proposed in Ref. \cite{RC}
since it allows one to increase the success rate to hit the searched state.

The paper is organized as follows. In Sec. II, we describe the
model and the strategies to achieve the quantum search. In particular, the
local and parallel strategies are presented. Section III is
devoted to the definition of a cost and the calculation of the
non-adiabatic losses which are used to define the optimality and to compare the local and parallel
strategies. The comparison is illustrated numerically in Sec.
IV while the conclusions are given in Sec. V.

\section{Model and strategies}
The marked state  $|m \rangle$ is one of the computational basis
states $|1\rangle,\cdots,|N\rangle$. The two-parameter Hamiltonian
(\ref{H2}) can be rewritten in an orthogonal basis which
features the marked state $|m \rangle$ and the uniform
superposition of unmarked states, $|u\rangle=\sum_{i \neq m }
\frac{1}{\sqrt{N-1}} | i \rangle$,
\begin{eqnarray}
\label{H2}
H&=& \left( \frac{a+b}{2}-\Delta \right) |m\rangle \langle m|+\left( \frac{a+b}{2}+\Delta \right) |u\rangle \langle u|\NN\\
&+&\Omega \left(|u\rangle \langle m|+|m\rangle \langle u| \right),
\end{eqnarray}
%\begin{eqnarray}
%\label{H2}
%H&=&\left[\begin{array}{cc} a(1-1/N) & a\sqrt{N-1}/N\\
%a\sqrt{N-1}/N & b+a/N\end{array}\right]\cr
%&=&\frac{a+b}{2}\openone_2+
%\left[\begin{array}{cc} \Delta & \Omega\\
%\Omega & -\Delta\end{array}\right]
%\end{eqnarray}
with $\Omega=a\sqrt{N-1}/N$ and $\Delta=(a-b)/2-a/N$.
%We diagonalize $H$ using the transformation
%\BE
%R= \cos\theta  \left( |u\rangle \langle u|-  |m\rangle \langle m|  \right)+ \sin\theta \left(|w\rangle \langle m|+|m\rangle \langle w| \right)
%\EE
%\begin{equation}
%R=\left[\begin{array}{cc} \cos\theta & \sin\theta\\
%\sin\theta & -\cos \theta\end{array}\right], \quad
%\tan 2 \theta=\frac{\Omega}{\Delta},\quad \theta\in[0,\pi/2[
%\end{equation}
Its eigenvalues are
\begin{equation}
\label{vp}
\lambda_\pm=\frac{a+b}{2}\pm \frac{1}{2}\sqrt{a^2+b^2-2ab\left(1-\frac{2}{N}\right)},
\end{equation}
%\begin{equation}
%\psi_+=\left[\begin{array}{c} \cos\theta\\
%\sin\theta \end{array}\right],\quad
%\psi_-=\left[\begin{array}{c} \sin\theta\\
%-\cos \theta\end{array}\right],
%\end{equation}
and the pertaining  eigenvectors read
\BEA
|+\rangle& &= \cos\theta | u \rangle+\sin\theta  | m \rangle, \quad \tan 2 \theta=\frac{\Omega}{\Delta},\quad \theta\in[0,\pi/2[\NN\\
|-\rangle& &= \sin\theta | u \rangle-\cos\theta  | m \rangle .
\EEA

In the adiabatic representation, the Hamiltonian becomes
\BEA
 \label{Hadiab}
H_{\rm ad}&=&
B^\dagger H B - i B ^\dagger \dot B \\
&=& \lambda_+ |+\rangle \langle +|+ \lambda_- |-\rangle \langle-|+i \dot \theta \left(|-\rangle \langle +|-|+\rangle \langle -| \right), \NN
\EEA
where the unitary transformation $B$ is formed by the instantaneous eigenstates of $H$, and with the off-diagonal non-adiabatic coupling
\begin{equation}
\label{thetadot}
\dot\theta=\frac{1}{2}%
\frac{\dot{\Omega}\Delta-\Omega\dot{\Delta}}{%
\Delta ^{2}+\Omega ^{2}}=\frac{\sqrt{N-1}}{N}%
\frac{a\dot{b}-\dot{a}b}{(\lambda_+-\lambda_-)^2}.
\end{equation}
In the adiabatic limit, when the characteristic time of the
process becomes arbitrarily large, the non-adiabatic coupling
$\dot\theta$ can be neglected and the dynamics follows the
adiabatic state(s) connected to the initial state.
For the search problem, the initial state is the uniform superposition
$|w\rangle=\frac{1}{\sqrt{N}} \sum_{i=1}^N | i \rangle$ which gives no particular role to any state of the computational basis.
Since $|w\rangle$ is the eigenvector of unit eigenvalue of $H_{\rm
i}$, taking $b(t_{\rm i})=0$ implies that  the
instantaneous eigenvector $|+\rangle$ is connected to $|w\rangle$ at the initial
time $t_{\rm i}$:
\begin{equation} \label{ci}
b(t_{\rm i})=0 \rightarrow |+\rangle (t_{\rm i})=|w\rangle.
\end{equation}
At the final time $t_{\rm f}$, we
require this eigenvector of higher eigenvalue to coincide  with the marked
target state, which is satisfied for $a(t_{\rm f})=0$,
\begin{equation}
\label{cf} a(t_{\rm f})=0 \rightarrow  |+\rangle(t_{\rm
f})=|m\rangle.
\end{equation}
The adiabatic theorem \cite{adiabatic1} can be recovered from (\ref{Hadiab}): starting from an instantaneous eigenvector, its
population remains larger than $1-\ep^2$ provided the ratio of the
off-diagonal coupling and the gap between the eigenvalues is at
least $\ep$, \BE \label{Tadiab} {\rm max}_{t \in [t_{\rm i} ,
t_{\rm f}]}  \ \dot\theta <\varepsilon \  {\rm min}_{t \in [t_{\rm
i} , t_{\rm f}]} \  \frac{\lambda_+-\lambda_-}{2}. \EE

\subsection{Linear strategy}
A naive algorithm would interpolate linearly between the values of
$a$ and $b$ at the initial and final times, \BEA
a(t)&=&\alpha \frac{t_{\rm f}-t}{T_{\rm linear} }\NN\\
b(t)&=&\alpha \frac{t-t_{\rm i}}{T_{\rm linear} }, \EEA with
$\alpha$ some multiplicative constant which fixes the energy
levels and $T_{\rm linear} \equiv t_{\rm f}-t_{\rm i}$ the total
duration of the process. From (\ref{vp}) one deduces that the
smallest gap is $\alpha/\sqrt{N}$ while (\ref{thetadot}) yields
${\rm max}_t \  \dot\theta=\sqrt{N}/T_{\rm linear} $. The
adiabaticity condition (\ref{Tadiab}) thus implies that the
computational cost is of order $N$, \BE \label{T_naive}
 \alpha T_{\rm linear}   > 2\frac{N}{ \ep}.
\EE A quantum algorithm with such a linear dynamics therefore does
not perform better than a classical search. As noted in Ref.
\cite{RC}, this stems from the fact that by applying
(\ref{Tadiab}) globally, i. e., to the entire time interval, one
imposes a constraint on the evolution rate during the whole
computation while the constraint is only severe where the gap is
close to the minimum. In the next section, we recall the strategy
proposed by Roland and Cerf \cite{RC} which amounts to applying
locally the adiabatic theorem for infinitesimal time intervals and
adapting the rate at which the gap in eigenvalues is crossed. Our
approach, which is presented afterwards in Sec. \ref{seclevel},
consists in following level lines on the surface of eigenvalues
difference corresponding to parallel eigenvalues, i. e., instead
of following a given path at a varying speed, we follow a
different path at a constant speed. This approach has been shown
\cite{Leveline,GJ}, in a different context,  to both be robust and
strongly reduce the nonadiabatic losses.

\subsection{Local strategy}
The strategy proposed by Roland and Cerf \cite{RC} consists in applying  the adiabaticity condition
locally in time rather than on the whole interval as in (\ref{Tadiab})
and correspondingly adapting
the rate $\dot\theta$ at which the gap $\lambda_+-\lambda_-$ is crossed,
\begin{equation}
\dot\theta=\varepsilon \frac{\lambda_+-\lambda_-}{2}.
\label{thetadot_}
\end{equation}
This is equivalent to  fixing instantaneously the
non-adiabatic losses to $\ep^2$ at any time applying
time-dependent perturbation theory on the Hamiltonian (\ref{Hadiab}).
 With the
parametrization $a+b=\alpha$, (\ref{vp}) becomes \BE
\label{val_RC} \lambda_\pm=\frac{\alpha}{2}\pm \frac{\alpha}{2}
\sqrt{1-4 \frac{N-1}{N} a(1-a)}. \EE Hence, (\ref{thetadot_})
yields a differential equation for $a$, \BE \dot a=-\frac{\alpha
\ep}{2} \frac{N}{\sqrt{N-1}} \left[ 1-4 \frac{N-1}{N}
a(1-a)\right]^\frac{3}{2}. \EE Its implicit solution satisfying
the initial condition $a(t_{\rm i})=\alpha$, which arises because
of the requirement $b(t_{\rm i})=0$, reads
\begin{eqnarray}
\alpha(t-t_{\rm i}) =\frac{\sqrt{N-1}}{\ep}\left(1+\frac{1-2 a}{\lambda_+-\lambda_-}\right).
\label{anonlin}
\end{eqnarray}
At the final time, one has $a(t_{\rm f})=0$ so that, denoting the process duration by $T_{\rm local} = t_{\rm f} -t_{\rm i}$, one obtains
\begin{equation}
\label{T_RC}
\alpha T_{\rm local} =2\frac{\sqrt{N-1}}{\ep},
\end{equation}
which shows that for $\alpha$ of order $N^0$, the search duration
scales as $N^{1/2}$, in contrast to the linear strategy for which,
according to (\ref{T_naive}), it scales as $N$.

\begin{center}
\begin{figure}[h]
\includegraphics[scale=0.75]{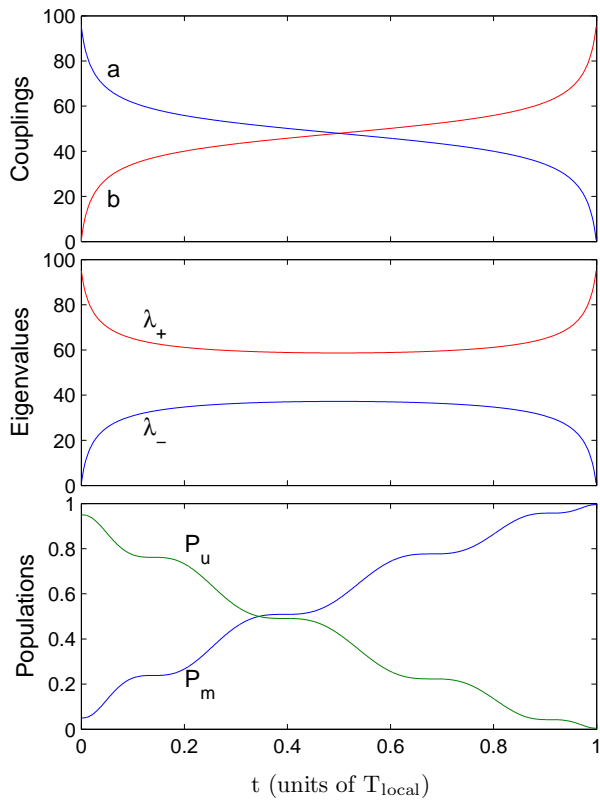}
\caption{(Color online) Dynamics of the local strategy search for $N=20$ and
$\ep=1/11$ (leading to $\alpha T_{\rm local}\approx 96$). Upper
frame: $a$ and $b$ (in units of $1/T_{\rm local}$); middle frame:
The eigenvalues $\lambda_{\pm}$ (in units of $1/T_{\rm local}$)
exhibiting an avoided crossing; lower frame: Populations
$P_u=|\langle u|\phi\rangle|^2$ and $P_m=|\langle
m|\phi\rangle|^2$. The search is achieved with probability
0.995.} \label{Dyn_loc}
\end{figure}
\end{center}

The inversion of (\ref{anonlin}) yields \BE \label{aloc}
a(t)=\frac{\alpha}{2}\left(1-
\frac{1}{\sqrt{N}}\frac{s(t)}{\sqrt{1-\frac{N-1}{N}s^2(t)}}\right),
\EE with $s(t)=\frac{2t-t_{\rm i}-t_{\rm f}}{T_{\rm local}} $.
%We shall denote by
%$T_{\rm loc}$ the process duration $t_f-t_i$.
The gap $\lambda_+-\lambda_-$ reads \BEA \label{diff}
\lambda_+-\lambda_-= \frac{\alpha}{\sqrt{N}} \frac{1}{\sqrt{1-
\frac{N-1}{N} s^2(t)}}, \EEA and its minimum is
$\frac{\alpha}{\sqrt{N}}$. Figure \ref{Dyn_loc} depicts an example
of the dynamics of the search for a high success rate.

\begin{center}
\begin{figure}[h]
\includegraphics[scale=0.75]{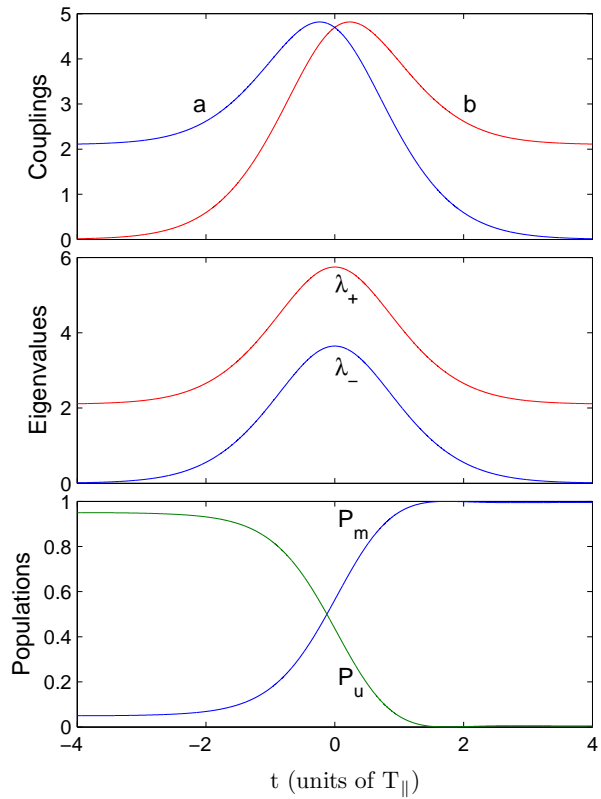}
\caption{(Color online) Dynamics of the parallel strategy search for
$F_{T_{\parallel}}(t)=\tanh(t/T_{\parallel})$, $N=20$ and $\beta
T_{\parallel}=4.7$. Upper frame: $a$ and $b$ (in units of
$1/T_{\parallel}$); middle frame: The parallel eigenvalues
$\lambda_{\pm}$ (in units of $1/T_{\parallel}$); lower frame:
Populations $P_u=|\langle u|\phi\rangle|^2$ and $P_m=|\langle
m|\phi\rangle|^2$. The search is achieved with probability
0.995.}\label{Dyn_para}
\end{figure}
\end{center}

\subsection{Parallel strategy}
\label{seclevel} The strategy we propose here consists in
following an appropriate level line on the surface of eigenvalues
difference as a function of the parameters $a$ and $b$ of the
Hamiltonian (\ref{H2}), corresponding to parallel eigenvalues. Let
$\frac{2\beta}{\sqrt{N}}$ denote this difference where $\beta$ is
some constant to be chosen while, as we shall see below, the
$\sqrt{N}$ arises to avoid energy blow up with $N$. From Eq.
(\ref{vp}), the level line
$\lambda_+-\lambda_-=\frac{2\beta}{\sqrt{N}}$ is given by the
ellipse
\begin{equation}
a^2+b^2-2ab\left(1-\frac{2}{N}\right)=\frac{ \beta^2}{N},
\end{equation}
or, in canonical form, \BE
\frac{(a+b)^2}{4\beta^2}+\frac{(b-a)^2}{\frac{4\beta^2}{N-1}}=1.
\label{ellipsecan} \EE
The initial condition (\ref{ci}), i. e.   $b(t_{\rm i})=0$,
implied that $a(t_{\rm i})=\frac{2\beta}{\sqrt{N}}$. At the final time
$t_{\rm f}$,  (\ref{cf}) holds so that
$b(t_{\rm f})=\frac{2\beta}{\sqrt{N}}$.
%Notice that the maximum of $a+b$
%is $\eta$:
%\begin{equation}
%\max(a+b)=\eta.
%\end{equation}
It follows that the parametric equation of the ellipse is
\BEA
a(t)&=&\beta \left(\sqrt{1-\frac{N-1}{N} F^2(t)}-\frac{F(t)}{\sqrt{N}}\right) \NN \\
b(t)&=&\beta \left(\sqrt{1-\frac{N-1}{N}
F^2(t)}+\frac{F(t)}{\sqrt{N}}\right) , \label{ab} \EEA with $F(t)$
a strictly monotonous function such that $F(t_{\rm i})=-1$ and
$F(t_{\rm f})=1$. Here we consider explicitly the case of the
(analytic) hyperbolic tangent of characteristic width
$T_{\parallel}$, $F(t)=\tanh(t/T_{\parallel})$. The eigenvalues
(\ref{vp}) read here \BE \lambda_\pm=
\beta\left(\sqrt{1-\frac{N-1}{N}
F^2(t)}\pm\frac{1}{\sqrt{N}}\right). \EE Figure \ref{Dyn_para}
displays an example of the dynamics of the search, illustrating
how the population transfer is achieved from the prepared state to
the marked one. In the numerics, we have truncated the time
interval from $t=-4T_{\parallel}$ to $t=4T_{\parallel}$. We have
checked that taking a larger interval does not change
significantly the result. We have chosen a situation leading to
the same efficiency as the local strategy example shown in Fig.
\ref{Dyn_loc}. 
%We remark however that the area of the coupling
%$a(t)$ is approximately three times larger for the local strategy
%than for the parallel one.

Comparing Figs. \ref{Dyn_loc} and \ref{Dyn_para}, one also notices
that the search is achieved in an oscillatory manner
for the local strategy whereas it is achieved in a monotonic manner
for the parallel strategy. This feature, as well as the
enhancement of the success rate for the parallel strategy, can be
interpreted using superadiabatic basis that are better adapted to
describe the dynamics \cite{Berry,Joye}. In
general, for a strategy using  analytic coupling parameters (as considered in the
parallel strategy), in the adiabatic limit, the non-adiabatic
losses are exponentially small, i.e. of the form  $\sim
e^{-|\text{const.}|T_{\parallel}}$ and thus beyond any power of
$1/T_{\parallel}$ while the corresponding history of the dynamics
is smooth and monotonic. For the local strategy, the condition
(\ref{thetadot_}) prevents the analyticity of the
parameters. 
The discontinuity of the coupling  gives rise to losses which are of order 2 in the off-diagonal elements of (\ref{Hadiab}).
More generally,  a discontinuity of the $n^{\hbox{\scriptsize th}}$
derivative of the couplings corresponds to polynomial non-adiabatic
losses of the order of  $(1/T^{n+1})^2$, and the corresponding
dynamics exhibits oscillations. Indeed, the transformation leading to (\ref{Hadiab}) can be iterated, that is one may further diagonalize (\ref{Hadiab}), which  yields higher order derivatives \cite{Joye,Drese}.

As we show more precisely in the next section, for an identical
search cost, the parallel strategy generally enhances the success
rate, i. e. reduces the non-adiabatic losses, with respect to the
local strategy, or equivalently reduces the cost for an identical
success rate.

% and the
%parameters $a$ and $b$ as well as the instantaneous eigenvalues as
%a function of time. It also.

\section{Comparison: cost and losses}
In order to compare the local and parallel strategies, we shall
take into account both the computational cost and the success rate
of the search.
\subsection{Search cost} \label{sec_cost}
In actual implementations, the time-dependent coupling parameters
$a(t)$ and $b(t)$ can be achieved, for instance, by laser fields
\cite{cavity}. As a measure of the cost needed to achieve the
Grover search we can consider a quantity which is the equivalent
of the total laser power. Note that the coupling parameters can
either vary significantly during the whole duration of the process
(e. g. in the local strategy) or during a small fraction only (e.
g. in the parallel strategy). In order to account for both
situations, we define the cost $C$ as the product of the peak
value $a_{\rm peak}$ of the coupling parameter $a(t)$ and the
effective duration of the search $T_{\rm eff}$ \BE C=a_{\rm peak}
T_{\rm eff}, \label{Cpeak} \EE where $T_{\rm eff}$ is directly
related to the characteristic time $T$ of variation of $a(t)$ and
is typically a few repetitions of it, $T _{\rm eff}=r T$. Indeed,
considering $a(t)$ as a pulse, $T$ is its characteristic width
whereas $T_{\rm eff}$ is the full duration for which $a(t)$ is
significantly different from its asymptotic $t \rightarrow \pm
\infty$ values (hence it can be defined rigorously given some
tolerance level). Note that one could define the cost as the area
under $a(t)$ for this effective duration. The result would
generally differ only by a numerical factor close to one whereas
(\ref{Cpeak})
 is usually more convenient to compute.

For the local strategy, one deduces from (\ref{aloc}) that the
effective duration $T_{\rm eff}$ corresponds to the whole duration
$T_{\rm local}=t_{\rm f}-t_{\rm i}$ given in (\ref{T_RC}). The
cost reads thus \BE C_{\rm local}=\alpha T_{\rm local} =\frac{2
\sqrt{N-1}}{\ep}. \EE

For the parallel strategy, we consider analytical functions such
as $F(t)=\tanh(t/T_{\parallel})$, which approach their asymptotic
values for times $t$ with $|t|  \gtrsim r  T_{\parallel} /2$. From
(\ref{ab}), one obtains the peak value $a_{\rm peak}=\frac{\beta
(N-2)}{\sqrt{N(N-1)}}$. Hence, we have \BE
C_{\parallel}=\frac{\beta (N-2)}{\sqrt{N(N-1)}} r
T_{{\parallel}}\sim \beta r  T_{{\parallel}}. \EE Note that this
relation seems to imply that $T_{{\parallel}}$ can be choosen
arbitrarily (for instance of order $N^0$); However, the
non-adiabatic losses, studied below, would then increase
dramatically.

We shall compare the two strategies at identical costs and then
focus our attention below on the success rate of the search.
Requiring an identical cost for both strategies yields \BE
\label{cost2} \beta r T_{{\parallel}} =\frac{2 (N-1)\sqrt{N}}{
(N-2)      \ep} \sim \alpha T_{\rm local} . \EE For large $N$, the
characteristic time of the squared hyperbolic secant, associated
with the same cost for both strategies,  is thus just $(\alpha /
\beta r) \times T_{\rm {local}} $. Moreover, we can require the
effective duration of the search to be equal for both strategies
which simply amounts to having $\alpha=\beta$. Note that if one
chooses $\alpha=\beta=1$, then the search cost is directly equal
to the effective search duration. Since there is no  loss of
generality, we shall assume $\alpha=\beta=1$ throughout.
%Taking into account that one typically requires a few characteristic times
%to contain most of the pulse (i. e. the part of the pulse which differs significantly from 0),
%we shall take $\eta/2$ equal to that number (and take it equal to 8 in the examples).

\subsection{Non-adiabatic losses}
The success rate of the search is determined by the probability to
find the system in the marked state at the final time. This is
given by $1-P_{\text{loss}}$ where $P_{\text{loss}}$ corresponds
to the probability of the non-adiabatic losses from the
instantaneous  eigenstate initially populated to the other states.
These losses arise because the adiabatic state connected to the
initial state is not strictly followed as $\dot\theta$ is not
strictly zero. For the local strategy, the loss at the final time
can be calculated exactly for any $N$. Indeed, the Hamiltonian
(\ref{Hadiab}) can be rewritten as
\begin{widetext}
\BEA
\label{Hadiab2}
H_{\rm ad}= |+\rangle \langle +|+  |-\rangle \langle-| + \frac{\lambda_+-\lambda_-}{2}
 \left\{    |+\rangle \langle +|-  |-\rangle \langle-|+ i \ep
 \left(|-\rangle \langle +|-|+\rangle \langle -| \right) \right\},
\EEA
%
%\BEA
%\left[
%\begin{array}{cc}
%\lambda_+ & -i\dot\theta \\
%i\dot\theta & \lambda_-
%\end{array}
%\right] =\frac{\lambda_++\lambda_-}{2}\openone_2+\frac{\lambda_+-\lambda_-}{2}
%\left[
%\begin{array}{cc}
%1 & -i\ep \\
%i\ep & 1
%\end{array}
%\right] ,
%\EEA
where $\lambda_+-\lambda_-$ depends on time according to (\ref{diff}), and use was made
of (\ref{val_RC}) to get $\lambda_++\lambda_-=\alpha=1$.
Upon extracting the first term which  gives rise to a phase and defining a new time
 $ \tau(t) =\int^t \frac{\lambda_+-\lambda_-}{2} d t$  one obtains a stationary Hamiltonian.
It follows that, up to a phase, the state of the system at the new time $\tau$ is thus
\BEA
| \psi \rangle (\tau)=\frac{\ep}{\sqrt{1+\ep^2}} \sin (\sqrt{1+\ep^2} \tau)  |-\rangle
+\left( \cos (\sqrt{1+\ep^2} \tau)-\frac{i}{\sqrt{1+\ep^2}} \sin (\sqrt{1+\ep^2} \tau) \right )|
+ \rangle .
\EEA
\end{widetext}
From (\ref{val_RC}), we deduce that at the end of the process,
$\tau(t_{\rm f})=\frac{1}{\ep}\arctan \sqrt{N-1}$. The
non-adiabatic losses are therefore \BE \label{Lossloc}
P_{\text{loss,local}}=\frac{\ep^2}{1+\ep^2} \sin^2
\left(\frac{\sqrt{1+\ep^2}}{\ep} \arctan \sqrt{N-1} \right), \EE
i.e., in the limit of large $N$ and small $\ep$
 \BE P_{\text{loss,local}}\sim\ep^2\sin^2
\left(\frac{\pi}{2\ep}\right).
\EE
Note that the choice of the specific values
$\ep=1/2p$ with an integer $p$ will give losses going to 0 for
large $N$. However these choices of specific and thus non-robust
values will not be considered here since they are outside the
adiabatic scope. A good measure of the losses is the upper
boundary $\ep^2$.
The adiabatic regime for the local strategy is
thus reached when $\ep\ll1$, i.e., using (\ref{T_RC}),  when
\BE
\label{ad_RC}
\frac{T_{\text{local}}  }{\sqrt{N}} =\frac{2}{\ep}\gg1 \qquad   N \gg 1.
\EE

 For the level line optimization, it has been shown in Ref. \cite{Leveline}
 that the adiabatic regime is obtained when
$ T_{\parallel } (\lambda_+-\lambda_-)=2 T_{\parallel }
/\sqrt{N}\gg1$.
 Actually, for  $F(t)=\tanh(t/T_{\parallel})$, we can calculate
the non-adiabatic losses for large $N$,
%, i.e. when $\eta T_\parallel/\sqrt{N}\gg1$ and $N\gg1$:
\begin{equation}
\label{asympt} P_{\text{loss, $\parallel$}}\sim
\text{sech}^2\left(\frac{\pi T_{\parallel}\beta}{\sqrt{N}}\right)
.
%\sim e^{-\pi T_\parallel\eta/\sqrt{N}}.
\end{equation}
This result comes from the fact that, for $N\gg1$, the model
(\ref{H2}) corresponds, up to a phase, to the Allen-Eberly model
\cite{AE}. Equation (\ref{asympt}) shows the $\sqrt{N}$ scaling of
the search cost since taking $T_{\parallel}$ growing as
$\sqrt{N}$,
\begin{equation}
\label{ad_level}
\frac{T_{\parallel}}{\sqrt{N}}\equiv\frac{1}{\gamma} \gg 1 \qquad
N \gg 1,
\end{equation}
allows one to obtain the same arbitrarily
small non-adiabatic losses for any $N$,
\begin{equation}
\label{asympt2} P_{\text{loss,$\parallel$}}\sim
\text{sech}^2\left( \frac{\pi}{\gamma} \right)\sim 4 \exp\left(-
\frac{2 \pi}{\gamma} \right) .
%\sim e^{-\pi T_\parallel\eta/\sqrt{N}}.
\end{equation}

Requesting the same cost for both strategies, we deduce from
(\ref{cost2}), (\ref{ad_RC}) and (\ref{ad_level}) \BE
\label{loss2} \gamma\sim\frac{\ep r}{2}. \EE The losses
(\ref{asympt2}), being of the form $e^{-2 \pi/\gamma}$, are beyond
any power of $1/\gamma$, and are thus expected to be much smaller
than the ones given by the local strategy which are of order
$\ep^2$.
 %\BEA
%\dot{\theta_{\parallel}}&=&\frac{1}{2}\sqrt{\frac{N-1}{N}} \frac{\dot F}{\sqrt{1-\frac{N-1}{N}  F^2}}.
%\EEA
%The loss at the final time can be calculated exactly in the limit $N \rightarrow \infty$.

\section{Numerical illustration}
\begin{center}
\begin{figure}
\includegraphics[scale=0.75]{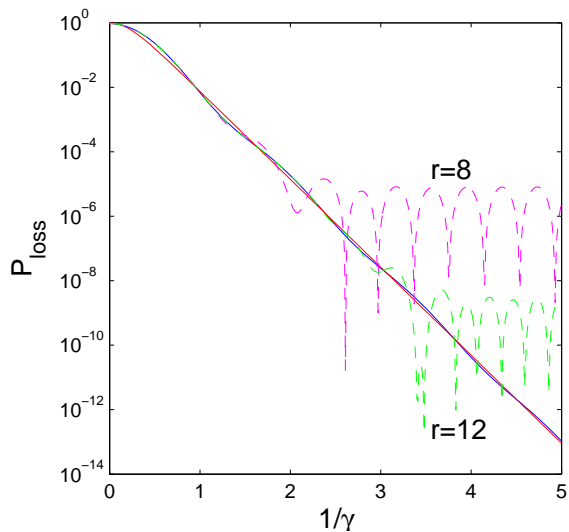}
\caption{(Color online) Final losses (in logarithmic scale) as a function of
$1/\gamma=T_{{\parallel}}/\sqrt{N}$ (dimensionless)  for $N=20$
with $F(t)=\tanh(t/T_{\parallel})$ (oscillating full line) its
asymptotic value given by (\ref{asympt}) (non-oscillating full
line, almost undistinguishable from the oscillating full line),
and for truncated time domains $r=12$ and $r=8$ (dashed lines).}
\label{Loss_tanh}
\end{figure}
\end{center}

The quadratic speedup of the search using the parallel strategy
was derived on the basis of the asymptotic result (\ref{asympt})
obtained for large $N$. We first  show that the losses are also
well approximated by (\ref{asympt}) for finite $N$. In Fig.
\ref{Loss_tanh} we plot the non-adiabatic losses obtained
numerically as a function of $1/\gamma$ defined as
$T_{\parallel}/\sqrt{N}$. The case $N=20$ (oscillating full line)
is close to the asymptotic result  (non-oscillating one) which, as
expected from  (\ref{asympt}), shows a strong exponential decay.
The search is efficient for $1/\gamma \ge 1$, with for instance
$P_{\text{loss, $\parallel$}}\sim10^{-2}$ for $\gamma=1$. Similar
results hold for other values of $N$.

In contrast to the local strategy, the parallel strategy uses
analytical couplings on an unbounded domain. Hence, in practice
one has to truncate this domain to limit the time of search. This
truncation of the couplings, breaking the continuity, leads in
general to additional non-adiabatic losses. In Sec.
\ref{sec_cost} we defined the finite domain through the quantity
$r$ as $T_{\text{eff}}=r T_{\parallel}$.
 Figure \ref{Loss_tanh} shows the losses for $r=8$ and $r=12$.
As expected, the loss becomes larger for smaller $r$ when
decreasing $\gamma$. For $r=12$, the range of
 validity of the asymptotic formula (\ref{asympt2}) is approximately $1/\gamma \lesssim 3$.
 This means that up to $1/\gamma \sim 3$ the additional losses due to the truncation can be neglected
(otherwise, one could take a larger value for $r$).

\begin{center}
\begin{figure}
\includegraphics[scale=0.75]{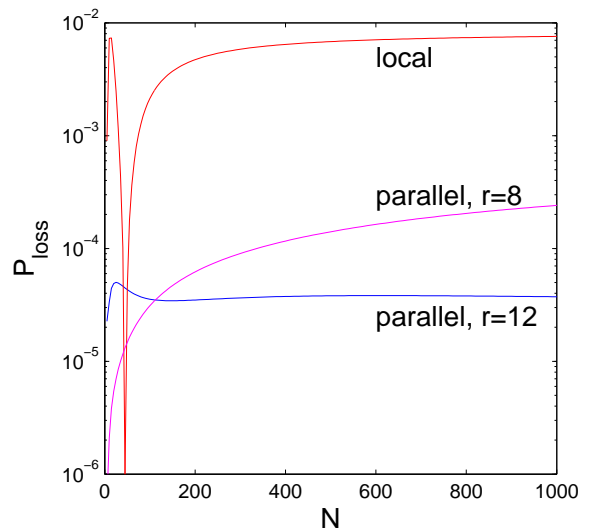}
\caption{(Color online) Final losses as a function of $N$ for $\varepsilon=1/11$
for the local strategy and for the parallel strategy with
$F(t)=\tanh(t/T_\parallel)$ with a truncated time
intervals ($r=12$ and $r=8$) and $\gamma$ given by Eq.
(\ref{loss2}) to have the same cost.} \label{Loss_tanh_N}
\end{figure}
\end{center}

Figure \ref{Loss_tanh_N} depicts the losses as a function of $N$
for a given value of $\ep$ and the corresponding value of $\gamma$ given by (\ref{loss2})
 with two truncations $r=8$ and $r=12$.
This figure shows that the parallel strategy is more efficient, as
expected from the comparison of (\ref{ad_RC}) and (\ref{ad_level}) with  (\ref{loss2}), despite the truncation of the time interval which
breaks the analyticity of the coupling. Note, however, that a truncation
with too small a value for $r$ can lead to a significant dependance of the losses with $N$.

\section{Conclusion}

We have proposed a strategy to achieve the Grover
search by adiabatic passage using parallel eigenvalues. We have
compared this parallel strategy with the known local strategy
which requires an adaptation of the speed of the dynamics with
respect to the given dynamical gap between the eigenvalues. We
have shown the superiority of the parallel strategy: for an
identical search cost, the parallel strategy enhances
the success rate  with
respect to the local strategy, i. e. reduces the non-adiabatic losses, or equivalently reduces the cost
for an identical success rate.

Smooth analytic coupling parameters are in principle
required for the parallel strategy. We have however shown
numerically that a truncation of the time domain, which is necessary in practice, preserves the
higher efficiency of the success rate of the search at identical cost.

We have here used an hyperbolic tangent for $F(t)$ since it allows one to determine analytically the population dynamics for large $N$. We have checked that other similar shapes for
$F(t)$, for instance associated to Gaussians which are easily performed in the laboratory, preserves the advantage of the parallel strategy over the local one.

\section*{Acknowledgments}
The authors are grateful to H.-R. Jauslin for useful discussions
and acknowledge support from the EU projects QAP and COMPAS, from the
Belgian government programme IUAP under grant V-18, and from the
Conseil R\'egional de Bourgogne. S.G. acknowledges support from
the French Agence Nationale de la Recherche (ANR CoMoC).


\begin{references}
\bibitem{adiabatic1} E. Farhi, J. Goldstone, S. Gutmann, and M. Sipser,
e-print quant-ph/0001106.
\bibitem{adiabatic2} A. M. Childs, E. Farhi, and J. Preskill,  Phys. Rev. A {\bf 65}, 012322 (2001).
\bibitem{nielsen} M. A. Nielsen and I. L. Chuang, {\em Quantum Computation and Quantum Information}, Cambridge University Press, Cambridge, 2000.
\bibitem{farhi} E. Farhi and S. Gutmann, Phys. Rev. A \textbf{57}, 2403 (1998).

\bibitem{RC} J. Roland and N. J. Cerf, Phys. Rev. A \textbf{65}, 042308 (2002).
%\bibitem{DDP} A. M. Dykhne, Sov. Phys. JETP {\bf 11}, 411 (1960);
%JETP 14, 941 (1962); J. P. Davis and P. Pechukas, J. Chem. Phys.
%{\bf 64}, 3129 (1976); A. Joye, H. Kuntz, and C.-Ed. Pfister, Ann.
%Phys. {\bf 208}, 299 (1991); A. Joye, G. Mileti, and C.-Ed.
%Pfister, Phys. Rev. A {\bf 44}, 4280 (1991).
%
%\bibitem{Super} M. V. Berry, Proc. R. Soc. London A
%{\bf 429}, 61 (1990); Proc. R. Soc. London A {\bf 414}, 31 (1987);
%A. Joye and C.-E. Pfister, J. Math. Phys. {\bf 34}, 454 (1993).

\bibitem{Leveline} S. Gu\'erin, S. Thomas and H. R. Jauslin,
Phys. Rev. A {\bf 65}, 023409 (2002).
\bibitem{GJ} S. Gu\'erin and H. R. Jauslin, Adv. Chem. Phys. {\bf 125}, 147 (2003).
%\bibitem{grover} L. K. Grover, Phys. Rev. Lett, {\bf 79}, 325 (1997).
%\bibitem{optics1} N. Bhattacharya, H. B. van Linden van den Heuvell, and R. J. C. Spreeuw, Phys. Rev. Lett. {\bf 88}, 137901 (2002).

\bibitem{Berry} M. V. Berry, Proc. Roy. Soc. Lond. A \textbf{429}, 61 (1990).
\bibitem{Joye} A. Joye, C.-E. Pfister, J. Math. Phys. \textbf{34}, 454
(1993).
\bibitem{Drese} K. Drese and M. Holthaus, Eur. Phys. J. D {\bf 3}, 73-86 (1998).
\bibitem{cavity} D. Daems and S. Gu\'erin, Phys. Rev. Lett. \textbf{99}, 170503 (2007); Phys. Rev. A {\bf 78}, 022330 (2008).
\bibitem{AE}  L. Allen and J. H. Eberly, {\em Optical Resonance and
Two-Level Atoms }(Dover, New York, 1987).
%\bibitem{Elk} M. Elk, Phys. Rev. A \textbf{52}, 4017 (1995).
%\bibitem{Drese} K. Drese and M. Holthaus, Eur. Phys. J. D \textbf{3}, 73 (1998).
%\bibitem{Kuhn} S. Nussmann, M. Hijlkema, B. Weber, F. Rohde, G. Rempe, and A. Kuhn,
%Phys. Rev. Lett. \textbf{95}, 173602 (2005).
%\bibitem{fstirap_exp} V.A. Sautenkov, C.Y. Ye, Y.V. Rostovtsev, G.R. Welch, and M. O. Scully,
%Phys. Rev. A \textbf{70}, 033406 (2004).
\end{references}
\end{document}